\documentclass[final,5p,times,twocolumn,fleqn]{elsarticle}

\usepackage{amssymb}

\usepackage[labelfont=bf,justification=raggedright,singlelinecheck=false]{caption}
\usepackage[utf8]{inputenc}

\usepackage{amsmath}
\usepackage{dcolumn}
\usepackage{bm}
\usepackage[colorlinks=true,allcolors=blue]{hyperref}
\usepackage{color}
\usepackage[dvipsnames]{xcolor}

\journal{Physica D}

\begin{document}

\begin{frontmatter}



\title{Measure, dimension, and complexity of the transient motion in Hamiltonian systems}


\author[]{Vitor M. de Oliveira\corref{cor1}}
\ead{oliveira.vitormartins@gmail.com}
\author[]{Matheus S. Palmero}
\author[]{Iberê L. Caldas}

\address{Institute of Physics, University of São Paulo, São Paulo, SP, 05508-090, Brazil}

\cortext[cor1]{Corresponding author}


\begin{abstract}

Hamiltonian systems that are either open, leaking, or contain holes in the phase space possess solutions that eventually escape the system's domain. The motion described by such escape orbits before crossing the escape threshold can be understood as a transient behavior. In this work, we introduce a numerical method to visually illustrate and quantify the transient motion in Hamiltonian systems based on the \emph{transient measure}, a finite-time version of the natural measure. We apply this method to two physical systems: the single-null divertor tokamak, described by a symplectic map; and the Earth-Moon system, as modeled by the planar circular restricted three-body problem. Our results portray how different locations for the ensemble of initial conditions may lead to different transient dynamical scenarios in both systems. We show that these scenarios can be properly quantified from a geometrical aspect, the \emph{transient correlation dimension}, and a dynamical aspect, the \emph{transient complexity coefficient}.

\end{abstract}




\end{frontmatter}


\section{Introduction}

Near-integrable Hamiltonian systems are characterized by a rich dynamical setting with the usual presence of both chaotic and regular motion when the system is under a small volume-preserving perturbation \cite{Lichtenberg1992}. In this case, its phase space is said to be \emph{mixed} as it is composed of regions of stability along with a chaotic sea.
In closed systems, chaotic orbits densely fill the chaotic area as the system is topologically transitive \cite{Devaney1989}. However, such motion is not uniform throughout the phase space and these orbits may temporarily concentrate in certain regions, a phenomenon called \emph{stickiness} \cite{Contopoulos2010}.

In open Hamiltonian systems, on the other hand, the transitivity property does not hold and hence distinct chaotic orbits may describe very different paths before escaping \cite{deOliveira2020}. In this situation, stickiness leads to \emph{dynamical trapping} since the \emph{escape orbits}, i.e., chaotic orbits that eventually exit the system's domain, stay in sticky regions for a considerable amount of time \cite{Zaslavsky1985}. Another property evident in open Hamiltonian systems is the role of the unstable and stable manifolds as the geometrical structures behind the system's dynamics.
For example, when there is more than one escape channel, there exist fractal boundaries between the escape basins corresponding to each exit. These boundaries are formed by invariant manifolds associated with certain unstable periodic orbits in the system \cite{Bleher1988}.

The dynamics of an escape orbit before reaching the escape threshold in an open system may be understood as a transient behavior. Leaking systems also present the same situation, with the leak corresponding to an escape condition in the phase space \cite{Altmann2013}. In this work, we address the transient dynamics of Hamiltonian systems and how it is affected by the choice of initial conditions. Specifically, we investigate how the paths described in the phase space by an ensemble of solutions prior to exiting the system differ from the paths taken by other ensembles. With this, we can assess which escape orbits experience dynamical effects such as stickiness and visually illustrate the influence of the system's underlying geometrical structures.
Such analysis is important for understanding the transient dynamics of various physical systems, especially in the fields of Plasma Physics and Celestial Mechanics, both of which are hallmarks of Hamiltonian mechanics \cite{Altmann2013, Abdullaev2014, Szebehely1967, Lai2011}.

We focus our analysis on two physical systems. The first one concerns the configuration of magnetic field lines in a single-null divertor tokamak, which is described by a symplectic map with one degree of freedom. The second one concerns the motion of a body with negligible mass under the gravitational influence of the Earth and the Moon, as modeled by the two-degrees-of-freedom planar circular restricted three-body problem. In both these systems, given our chosen parameters, we have a situation where there is only one exit and, hence, all the escape orbits belong to the same escape basin. 

For our investigation, we define a finite-time version of the natural measure specific for escape orbits: the \emph{transient measure}. By calculating this measure for an ensemble of initial conditions, we depict the transient motion associated with such ensemble on a given area of the phase space. Later, we characterize each case by defining two parameters: the transient correlation dimension, which is similar to the correlation dimension; and the transient complexity coefficient, which attributes weights to the ensembles based on particular dynamical properties. Our results show that both parameters are able to recognize and distinguish the complexity of each ensemble's dynamics.





In the literature on Hamiltonian systems, it is common knowledge that orbits that begin in distinct regions of the phase space may experience distinct stickiness effects \cite{Zaslavsky2007}. In that sense, the method outlined in this work offers a visual aid, along with a quantitative characterization, for the paths taken by the different ensembles. The detailed knowledge of these possible paths, namely what a given orbit may experience in this transient dynamics before escaping, is important and have further implications to both physical systems that are analyzed here. In magnetically confined plasmas, particularly in tokamaks assembled with a poloidal divertor, it was shown that the heat flux on the divertor plate closely follows the invariant manifolds associated with the unstable equilibrium created by the poloidal divertor \cite{Ciro2016}. Also, in Celestial Mechanics, for example, orbits in a thin chaotic layer along with a small dissipation may lead to the capture of irregular moons by giant planets \cite{Astakhov2003}.



This paper is organized as follows. In Sec.~\ref{sec:concepts} we define the mathematical framework used in this work, namely, the transient measure, the transient correlation dimension, and the transient complexity coefficient. In Sec.~\ref{sec:analysis} we apply our method to investigate the single-null divertor tokamak and the Earth-Moon system. Finally, we present our conclusions in Sec.~\ref{sec:conclusions}.

\section{Mathematical framework}
\label{sec:concepts}


\subsection{Mean transient measure}

Let ${\boldsymbol{\varphi}}_{t}(\boldsymbol{x}_0)$ be a solution of our dynamical system in the $D$-dimensional phase space with initial condition ${\boldsymbol{x}}_0$ and at time $t$, and let us cover the region of the phase space that we are interested in by a grid of $D$-dimensional boxes of side-length $\varepsilon$.
We call $\eta(B_i,{\boldsymbol{\varphi}}_{t}({\boldsymbol{x}}_0),T)$ the total time spent by the solution ${\boldsymbol{\varphi}}_{t}({\boldsymbol{x}}_0)$ inside the box $B_i$ in the time interval $t\in[0,T]$.

If $\eta$ is the same for almost every ${\boldsymbol{x}}_0$, the \emph{natural measure} for each box $B_i$ can be defined as \cite{Ott2002}
\begin{equation}
    \mu_i=\lim_{T\to\infty}\dfrac{\eta(B_i,{\boldsymbol{\varphi}}_{t}({\boldsymbol{x}}_0),T)}{T},
    \label{eq:natural_measure_def}
\end{equation}

\noindent if the limit exists.
It follows that $\sum_{i=1}^{N} \mu_i = 1$, where $N$ is the number of boxes in the grid, which depends on the box side-length $\varepsilon$.

The natural measure is defined in the asymptotic limit $T\to\infty$ and is usually associated with the dynamics of an orbit on a chaotic attractor. We are interested here, however, in the transient dynamics of escape orbits in Hamiltonian systems.
Then, we propose a finite-time version of Eq.~\eqref{eq:natural_measure_def}, which we call the \emph{transient measure},
\begin{equation}
    \nu_i=\dfrac{\eta(B_i,{\boldsymbol{\varphi}}_{t}({\boldsymbol{x}}_0),T^e)}{T^e},
    \label{eq:escape_measure_def}
\end{equation}

\noindent where $T^e$ is the escape time, i.e., the time it takes for the orbit that starts at ${\boldsymbol{x}}_0$ to reach a predefined escape region. Here, $\eta$ is the total time spent by the orbit inside the box $B_i$ before leaving the system. It is important to note that $\sum_{i=1}^{N} \nu_i = 1$.

If we consider an orbit in the chaotic sea, the transient measure reflects the path followed by the orbit up until exiting the system. Hence, this measure is able to depict the transient dynamics of an escape orbit, including effects such as stickiness.

An observation here is in order. In practice, we use $\tilde{T}^e=\min{(T^e,T_{max})}$ instead of $T^e$ in Eq.~\eqref{eq:escape_measure_def} since there is a computational time limit $T_{max}$ for which we can numerically integrate an orbit and it can be shorter than the orbit's escape time. Evidently, if $T_{max} > T^e$, then $\tilde{T}^e=T^e$. This point will be addressed further later in this section.


With the lack of the transitivity property, a chaotic orbit may escape before visiting all the available areas on phase space. Therefore, in order to better visualize the behavior of escape orbits, we define the \emph{mean transient measure}, the average of the transient measure on an ensemble $U$ composed by $M$ initial conditions,
\begin{equation}
    \bar{\nu}_i={\langle\nu_i\rangle}_{U}=\dfrac{1}{M}\sum^{M}_{j=1}\nu_{i,j},
    \label{eq:mean_escape_measure_def}
\end{equation}

\noindent where $\nu_{i,j}=\eta(B_i,{\boldsymbol{\varphi}}_{t}({\boldsymbol{x}}_{0,j}),T^e_j)/T^e_j$ is the transient measure for the $j$-th initial condition ${\boldsymbol{x}}_{0,j}$ and box $B_i$. As was the case for the transient measure, we have that $\sum_{i=1}^{N}\bar{\nu}_i=1$. 

Equation~\eqref{eq:mean_escape_measure_def} is well defined for any discrete ensemble. For us, $U$ has a small volume and a high number of elements $M$ which are uniformly distributed on a grid. The ensemble is centered at an initial condition of interest, and the mean transient measure, therefore, describes the transient dynamics associated with a small neighborhood of said point. In practice, $M$ is chosen high enough so that the orbits do visit a sufficient number of boxes and clearly depict the transient behavior in the phase space.

Apart from the finite-time aspect, another difference of Eq.~\eqref{eq:escape_measure_def} from Eq.~\eqref{eq:natural_measure_def} is that we do not demand it holds for almost every ${\boldsymbol{x}}_0$. With that, the mean transient measure, Eq.~\eqref{eq:mean_escape_measure_def}, is in fact a function of the ensemble of initial conditions:
\begin{equation}
    \bar{\nu}_i=\bar{\nu}_i(U).
    \label{eq:ensemble_dependency}
\end{equation}

Hence, there may be different transient behaviors in the system depending on the chosen ensemble of initial conditions $U$, which can be illustrated by calculating the mean transient measure profile. In this work, we are interested specifically in how the \emph{location} of $U$ affects the system's dynamics. Next, we present two approaches for quantifying Eq.~\eqref{eq:ensemble_dependency}.


\subsection{Transient correlation dimension}

The natural measure $\mu_i$ can also be seen as the visitation frequency on box $B_i$.
In the context of dissipative dynamical systems, this measure shows which boxes are more visited than others by a typical orbit on a chaotic attractor. Associated with such attractor, then, a spectrum of generalized dimensions $D_q$ for the continuous index $q$ can be defined as \cite{Ott2002}
\begin{equation}
    D_q=\dfrac{1}{1-q}\lim_{\varepsilon\to0}\dfrac{\ln I_q(\varepsilon)}{\ln (1/\varepsilon)},
    \label{eq:generalized_dimension}
\end{equation}

\noindent with
\begin{equation}
    I_q(\varepsilon)=\sum_{i=1}^{N_{V}(\varepsilon)}\mu_i^q,
    \label{eq:I_q}
\end{equation}

\noindent where $N_V\leq N$ is the number of visited grid boxes, which depends on the box side-length $\varepsilon$.


The main difference between the dimensions in Eq.~\eqref{eq:generalized_dimension} is given by Eq.~\eqref{eq:I_q}, which attributes weight to the visitation frequency depending on the value of $q$. Some $D_q$, such as $D_1$ for example, can be related to specific dynamical concepts \cite{Ott2002}. Here, we are interested in $D_2$, since $I_2$ scales in the same fashion as the correlation integral in a time series \cite{Grassberger1983}.

We can similarly define a spectrum of generalized dimensions associated with the mean transient measure, Eq.~\eqref{eq:mean_escape_measure_def}. However, instead of looking at the fractal geometry of an attractor, we are inspecting the paths taken by escape orbits in a Hamiltonian system. In special, we define the \emph{transient correlation dimension}, which is given by
\begin{equation}
    D_2^{{\bar{\nu}}}(U)=\lim_{\varepsilon\to0}\dfrac{\ln I_2^{\bar{\nu}}(\varepsilon)}{\ln \varepsilon},
    \label{eq:escape_correlation_dimension_def}
\end{equation}

\noindent with
\begin{equation}
    I_2^{\bar{\nu}}(\varepsilon)=\sum_{i=1}^{N_{V}(\varepsilon)}\bar{\nu}_i^2.
    \label{eq:I_2_escape}
\end{equation}

As the value of $D_2$ represents the degree to which the elements of an orbit are correlated, $D_2^{\bar{\nu}}$ returns similar information, but for an ensemble of escape orbits. We then expect the transient correlation dimension to be high when there is stickiness in the system, for instance, which makes it a suitable quantity for analyzing the transient behavior of the system.

For a $D$-dimensional phase space, we have that $D_2^{\bar{\nu}}\leq D$. In order to numerically determine $D_2^{\bar{\nu}}$, we first fix the ensemble $U$ and calculate $I_2^{\bar{\nu}}$ for different values of the box side-length $\varepsilon$. Later, we plot $I_2^{\bar{\nu}}$ as a function of $\varepsilon$ on a log-log plot and interpolate the result by means of a linear regression, as expressed by Eq.~\eqref{eq:escape_correlation_dimension_def}. The transient correlation dimension $D_2^{\bar{\nu}}$ is then given by the angular coefficient, i.e. the inclination, of the straight line representing the linear interpolation.


\subsection{Transient complexity coefficient}

We now introduce another quantity for characterizing the transient dynamics of escape orbits. Besides the visitation frequency, which is given by the mean transient measure, there are two other aspects that we can take into consideration for assessing the importance of a box on the grid.


First, the number of orbits $m_i$ that begin in the ensemble $U$ and pass through the box $B_i$ is usually not the same for all boxes. Hence, we may say that the grid boxes with higher values of $m_i$ have a higher influence on the ensemble and, consequently, on the system's transient properties.

Second, between the $m_i$ orbits that pass through a box $B_i$, the one with the largest escape time $T^e_j$ contributes the most to our analysis, since it reaches the highest number of box visitations. Therefore, we define $\tau_i=\max\limits_{j}\{T^e_j~|~  {\boldsymbol{\varphi}}_{t}({\boldsymbol{x}}_{0,j})\in B_i \ \text{for some} \ t<T^e_j\}$ and we consider a grid box more important if it has a higher $\tau_i$. Conversely, if all the orbits that go through a box $B_i$ rapidly exit the system, i.e. $\tau_i$ is small, we consider this box to be less important to the transient motion.

We then define the \emph{transient complexity coefficient} as
\begin{equation}
    c(U,\varepsilon) = \sum_{i=1}^{N}\alpha_i\beta_i\bar{\nu}_i,
    \label{eq:complexity_coefficient}
\end{equation}

\noindent where the weights of the boxes are given by
\begin{equation}
    \alpha_i=\dfrac{m_i}{M}
    \ \ \text{and} \ \ \beta_i=\dfrac{\tau_i}{T_{max}},
    \label{eq:complexity_alpha_beta}
\end{equation}

\noindent with $M$ and $T_{max}$, as introduced before, the number of initial conditions and the maximum integration time, respectively.

The coefficient $c$ gives an over-the-grid summation of the mean transient measure weighting in the two aforementioned aspects. While $\alpha_i$ reinforces the dependence on the ensemble, $\beta_i$ favors the system's slow dynamics. We also note that $c\leq1$, where $c=1$ in the improbable event that \emph{all} the orbits visit the same boxes with the \emph{same} escape time $\tau$ and $T_{max}=\tau$. The higher the value of the transient complexity is, the more complex are the paths taken by the orbits before leaving the system.

In practice, due to the time limitation $T_{max}$, only a subset  $\tilde{U}\subseteq U$ composed of $\tilde{M}$ initial conditions leads to trajectories that escape from the system. Therefore, since only escape orbits should contribute, we restrict the calculation of the transient complexity coefficient to $\tilde{U}$ and we use
\begin{equation}
    \tilde{\alpha}_i=\dfrac{\tilde{m}_i}{\tilde{M}}
    \ \ \text{and} \ \ \tilde{\beta}_i=\dfrac{\tilde{\tau}_i}{T_{max}}
    \label{eq:complexity_alpha_beta_tilde}
\end{equation}

\noindent in Eq.~\eqref{eq:complexity_coefficient} instead of $\alpha_i$ and $\beta_i$. Here, $\tilde{m}_i$ is the number of escape orbits that pass through box $B_i$ with initial condition in $\tilde{U}$ and $\tilde{\tau}_i$ is the longest escape time between these orbits.
If all orbits that begin in the ensemble $U$ escape, then $\tilde{T}^e=T^e$ for all orbits and Eqs.~\eqref{eq:complexity_alpha_beta_tilde} reduce to Eqs.~\eqref{eq:complexity_alpha_beta}.

As a last observation, we calculate the transient complexity coefficient for different ensembles in our analysis using the same box side-length $\varepsilon$. Hence, we can assume that the dependency on Eq.~\eqref{eq:complexity_coefficient} becomes $c=c(U)$ and use this quantity to compare the complexity of the transient motion between different cases.



\section{Transient motion analysis}
\label{sec:analysis}

In this section, we numerically investigate the transient behavior of escape orbits in two Hamiltonian physical systems: a tokamak equipped with a single-null poloidal divertor and the planar version of the Earth-Moon system.

For the tokamak system, escape orbits are related to the magnetic field lines that cross the poloidal divertor plate, carrying impurities and unwanted particles and, consequently, enhancing the tokamak performance. For the Earth-Moon system, escape orbits are the trajectories of small objects, such as artificial satellites and asteroids, which leave the Moon's realm of gravitational influence towards the Earth's vicinity.

For both systems, our defined grid does not cover the whole phase space, but rather a region $V$ in which we are interested. We, therefore, restrict our analysis to this region and consider only the total time spent inside the grid to calculate the transient measure, Eq.~\eqref{eq:escape_measure_def}. Furthermore, in order to deal with the practical limit on integration time, we choose a suitable $T_{max}$ to guarantee that, at least, $85\%$ of the orbits in an ensemble escape, i.e., $\tilde{M}\geq0.85M$.
By setting $T_{max}$ and $\tilde{M}$ large enough, we also assure that the calculated transient complexity coefficient, Eq.~\eqref{eq:complexity_coefficient}, is comparable between the different cases analyzed.


It is also important to note that the definitions presented in Sec.~\ref{sec:concepts} are based on grids formed by boxes with congruent sides. Therefore, we normalize the analyzed two-dimensional phase space of both systems to the unity square $[0,1]\times[0,1]$ when carrying out the numerical procedures. One can show that the phase space normalization does not interfere with the results obtained from Eqs.~\eqref{eq:natural_measure_def}-\eqref{eq:complexity_coefficient}. In particular, when calculating the transient correlation dimension, Eq.~\eqref{eq:escape_correlation_dimension_def}, the scaling changes the \emph{linear} coefficient of the $\ln{I_2^{\bar{\nu}}}\times\ln{\epsilon}$ graph, but not the \emph{angular} coefficient given by $D_2^{\bar{\nu}}$.

\subsection{Single-null divertor tokamak}

\begin{figure}[!ht]
\centering
\includegraphics[scale=0.40]{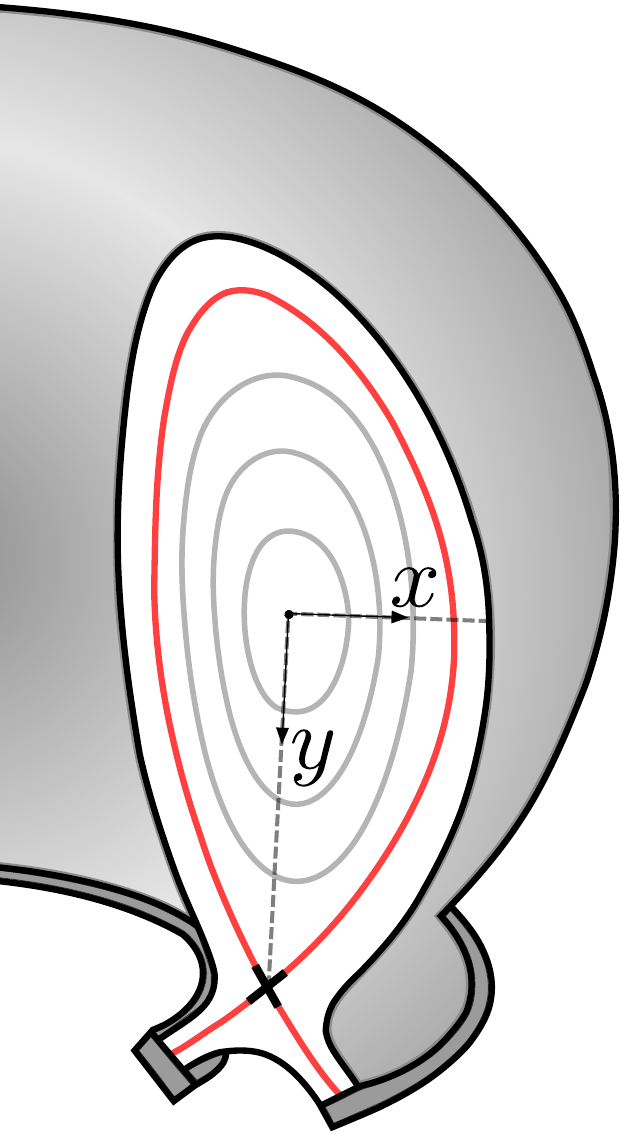}
\caption{Poloidal section of a divertor tokamak, showing the closed magnetic field lines (light gray lines), magnetic separatrix (red line), magnetic saddle (black cross) and the rectangular coordinates $(x,y)$.}
\label{fig:BM_schematics}
\end{figure}

Poloidal divertors are external magnetic coils that can be assembled in a tokamak\footnote{A tokamak is a toroidal shaped device that uses a strong magnetic field in order to confine a hot fusion plasma.} to conduct the magnetic field lines at the plasma edge towards an exit point.
Technically, the divertor induces a magnetic configuration with a single saddle point near the divertor plate known as the magnetic saddle. Due to perturbations on the magnetic configuration, a chaotic layer is formed around the saddle, allowing the magnetic field lines to escape this chaotic region through the divertor plate \cite{Kroetz2012}. Fig.~\ref{fig:BM_schematics} presents the system's schematic.

\begin{figure*}[!ht]
\centering
\includegraphics[scale=0.83]{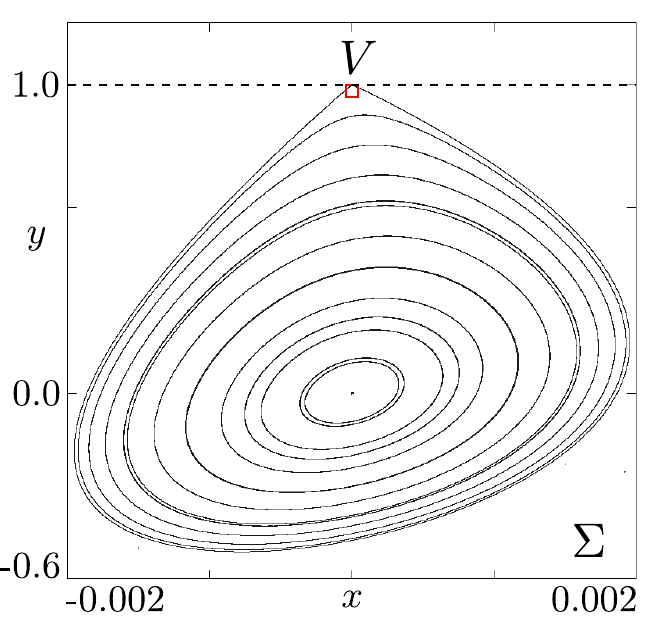}
\includegraphics[scale=0.83]{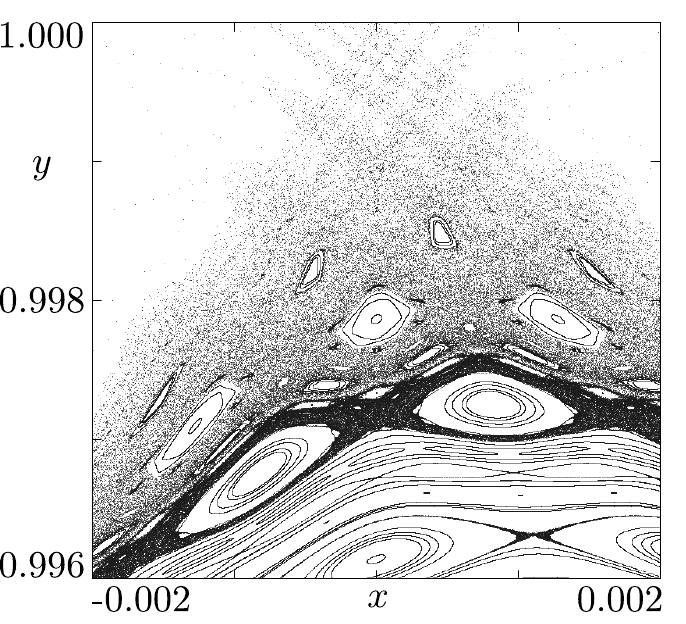}
\caption{Phase space $x$-$y$ of the single-null tokamak map. (Left) Full phase space, denoted as $\Sigma$, where the dashed line marks the escape threshold. (Right) Zoom-in on the $V$ region.}
\label{fig:BM_phase_space}
\end{figure*}

\begin{figure*}[!ht]
\centering
\includegraphics[scale=0.83]{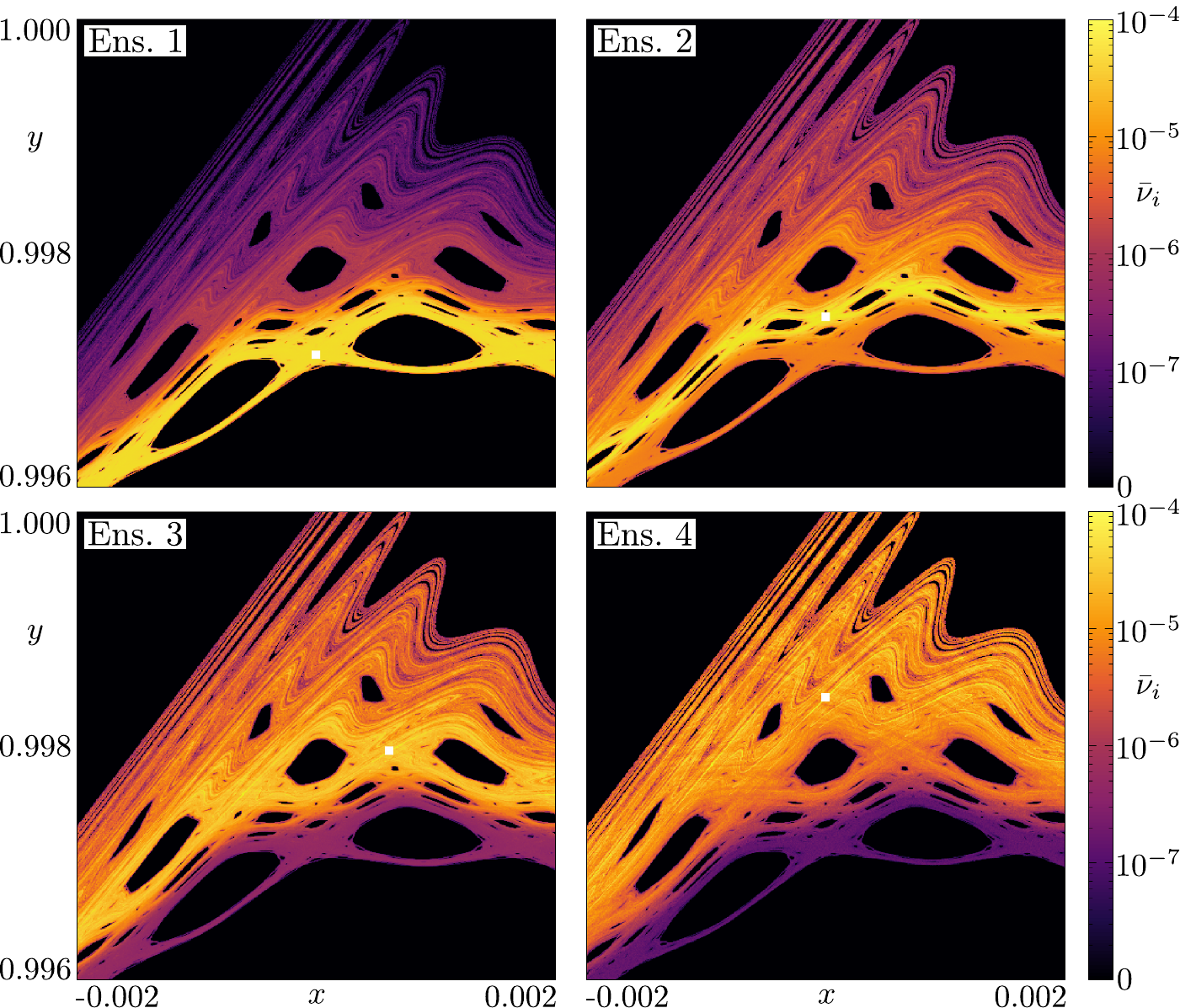}
\caption{Profiles of the mean transient measure $\bar{\nu}_i$ in logarithmic scale for the single-null tokamak divertor map, calculated on a $512\times512$ grid in the region $V$ of phase space $x$-$y$. The ensembles of initial conditions are represented by the small white squares which are not in scale.}
\label{fig:BM_profile}
\end{figure*}

\begin{figure*}[!ht]
\centering
\includegraphics[scale=0.83]{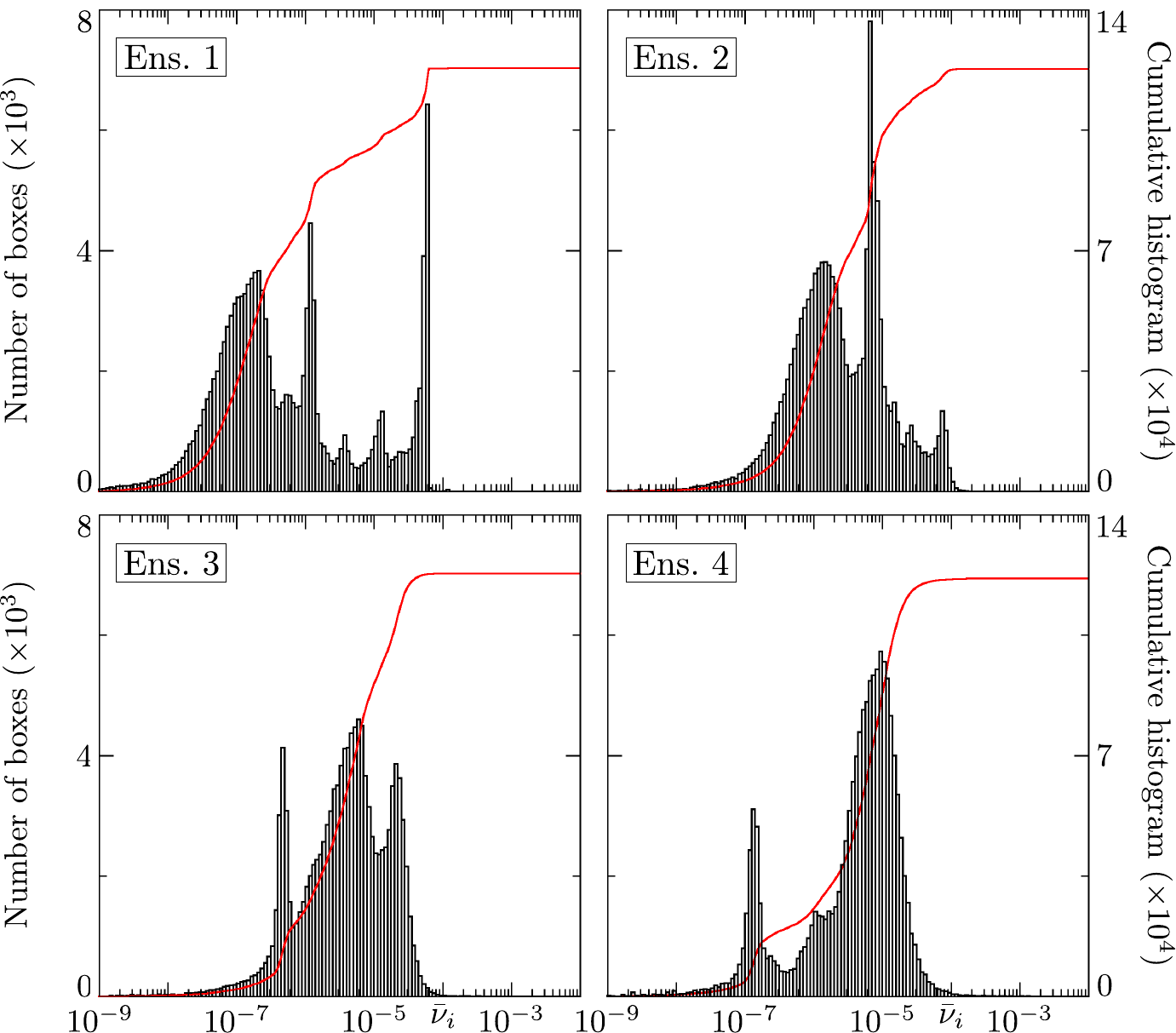}
\caption{Histogram (black rectangles) and cumulative histogram (red line) of the mean transient measure $\bar{\nu}_i$ for the single-null tokamak divertor map. The grid is formed by $512^2=26.2144\times10^{4}$ boxes and only the ones visited by an orbit are considered.}
\label{fig:BM_histogram}
\end{figure*}

\begin{figure*}[!ht]
\centering
\includegraphics[scale=0.83]{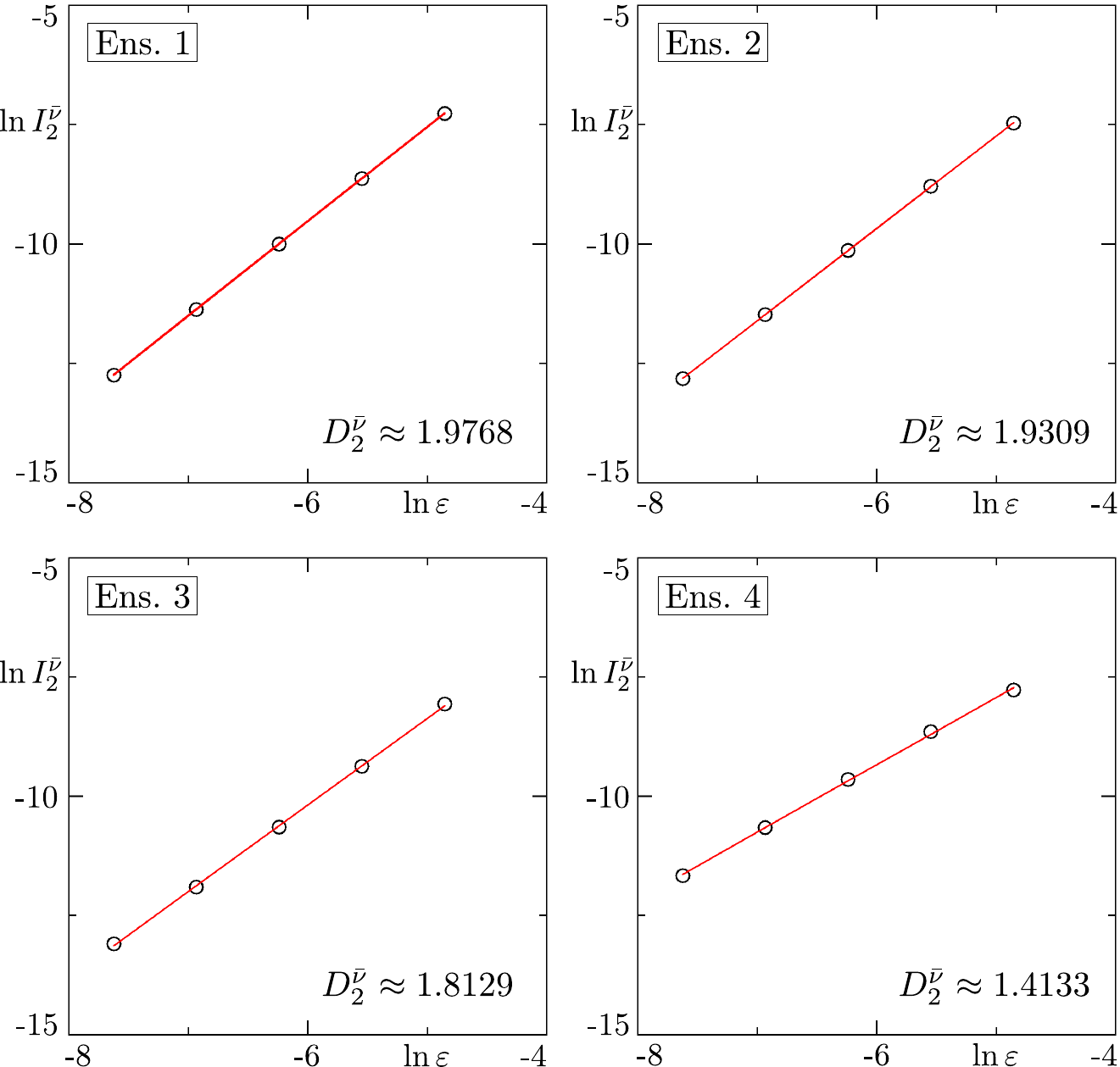}
\caption{Linear fitting of $\ln{I^{\bar{\nu}}_2}$ as a function of $\ln{\varepsilon}$ for the single null tokamak divertor map. $D^{\bar{\nu}}_2$ is the angular coefficient. The values of the box side-length are $\varepsilon=1/128$, $1/256$, $1/512$, $1/1024$ and $1/2048$ for a phase space normalized to the unity square.}
\label{fig:BM_dimension}
\end{figure*}

The symplectic map proposed in Ref.~\cite{Punjabi1992} is a phenomenological model for this system and it is given by
\begin{equation}
\begin{aligned}
x_{n+1}&=x_n-ky_n(1-y_n),\\
y_{n+1}&=y_n+kx_{n+1},\\
\end{aligned}
\label{eq:boozer_map}
\end{equation}

\noindent where $(x, y)$ are rectangular coordinates over a poloidal section surface, as depicted in Fig.~\ref{fig:BM_schematics}, and the control parameter $k$ is related to the amplitude of toroidal asymmetries that perturb the magnetic field configuration. Here, we use $k = 0.6$, which is adequate to simulate the diverted magnetic field configuration for large tokamaks like ITER \cite{Punjabi1997}.

The left panel of Fig.~\ref{fig:BM_phase_space} shows the system's phase space $x$-$y$. The magnetic saddle is located at $(x=0,~y=1)$ and we consider that a magnetic field line escapes when it crosses the divertor plate, i.e., the escape condition is given by $y>1$. We are interested in a sub-region $V$ which contains the saddle and is close to the escape threshold. The phase space in $V$ is presented in the right panel of Fig.~\ref{fig:BM_phase_space}. We note that the system possesses a separatrix chaotic layer embedded with several island chains.

For our numerical simulations, we choose four ensembles inside $V$ and we evolve them up to $T_{max}=2\times10^{6}$ iterations. Each ensemble is formed by a square of side-length $1 \times 10^{-5}$ and is composed of $M=10^4$ initial conditions uniformly distributed on a grid. In order to illustrate the different paths taken by the orbits in this system, we define a $512\times512$ grid and we calculate the mean transient measure profile for all cases. The results are shown in Fig.~\ref{fig:BM_profile}.

Since the phase space in $V$ is dominated by a complex configuration of island chains, it is reasonable that we place such ensembles on top of the unstable periodic orbits (UPOs) related to these islands. Ensembles 1 through 4 are then centered at UPOs of period 28, 57, 30 and 29, which are located at $(0.0,~0.9971)$, $(0.0,~0.9974)$, $(0.0006,~0.9979)$ and $(0.0,~0.9984)$, respectively.
We readily observe that the $\bar{\nu}_i$ profile in this region, as depicted by the logarithmic color scale, is highly dependent on the ensemble location as each case leads to different transient behavior. 
In especial, it highlights the stickiness experienced by the orbits that begin in Ens. 1 and 2.

In all cases, the color gradient depicts interesting structures formed between the island chains. These are actually the unstable manifolds associated with particular periodic orbits in the system and they are outlined by the trajectories as the discrete dynamics evolve \cite{Ciro2018}. We also notice the low values of the mean transient measure in the neighborhood of the islands, being especially visible for Ens. 4. This phenomenon is related to invariant manifolds as well or, specifically, to the distribution of heteroclinic crossings in the phase space \cite{deOliveira2020}.

To statistically investigate the mean transient measure profiles, we present both the histogram and the cumulative histogram for all four different ensembles in Fig.~\ref{fig:BM_histogram}. Here, we consider only the boxes visited at least once by the simulated dynamics. We quickly recognize that not only the histogram distributions but also the cumulative curves are quite different between the analyzed cases. Ensemble 1 shows a wider distribution in $\bar{\nu}_i$, presenting at least four distinct peaks. Meanwhile, Ens. 2, 3, and 4 show more centralized distributions, displaying a different number of peaks in each case, with the last one being the most well behaved.

The calculated histogram distributions stress the different transient behaviors which can emerge from the complex dynamical scenario of the system, as seen in Fig.~\ref{fig:BM_profile}.
All orbits beginning in Ens. 1 pass through all the island chains, following their invariant manifolds, before reaching the divertor plate at $y=1$. Ensemble 4, in particular, is located closer to the system's exit and the influence from the islands below it is low. 
Hence, the importance of the island chains regarding the paths followed by the escape orbits depends on the location of the chosen ensemble, which is translated as the number of peaks in Fig.~\ref{fig:BM_histogram}.

In order to quantify the differences illustrated by the mean transient measure profiles, we proceed with the calculation of the transient correlation dimension for each analyzed case. In Fig.~\ref{fig:BM_dimension}, we plot $I^{\bar{\nu}}_2$, Eq.~\eqref{eq:I_2_escape}, as a function of the box side-length $\varepsilon$. We see that all cases can be well fitted by a linear regression in the log-log plot, which corroborates  Eq.~\eqref{eq:escape_correlation_dimension_def}. Also, by comparing the calculated values for $D_2^{\bar{\nu}}$ to the profiles in Fig.~\ref{fig:BM_profile}, we find that the transient correlation dimension is well suited for characterizing the transient behavior in this system. As expected from Fig.~\ref{fig:BM_histogram}, these dimensions monotonically decrease as the ensemble of initial conditions gets closer to the escape threshold, which indicates that the transient behavior is more complex when the orbits begin far from the exit.

\begin{figure}[!ht]
\centering
\includegraphics[scale=0.83]{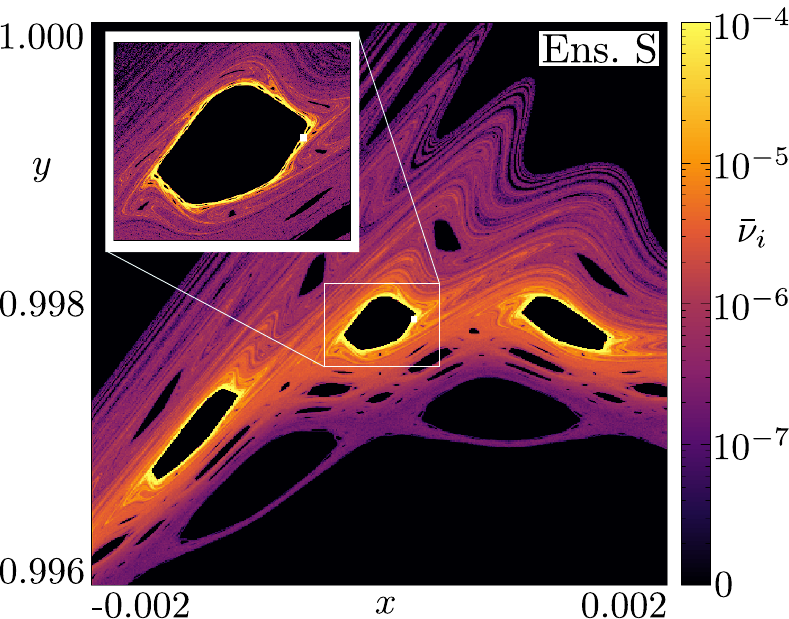}
\caption{Profile of the mean transient measure in logarithmic scale for the special case in the single null tokamak divertor map. On the inset, the same profile is calculated on a smaller region. The ensemble position is represented by a small white square in both figures.}
\label{fig:BM_special}
\end{figure}

We continue our analysis by considering a special case where we position an ensemble S in the neighborhood of a stability region. Specifically, S is centered at an UPO of period 464 associated with the satellite islands of the center island chain of period 30. Like the other cases, it is composed of $M=10^{4}$ initial conditions and we also iterate it until $T_{max}=2\times10^{6}$, but, this time, it is formed by a smaller square of side-length $1 \times 10^{-6}$. The mean transient measure profile for this special case is presented in Fig.~\ref{fig:BM_special} for a 512 $\times$ 512 grid. We consider the same region $V$ of the phase space that was used for the other ensembles and also a smaller region that focuses on the center island (inset). It is clear that the $\bar{\nu}_i$ profile highlights the presence of stickiness and outlines the invariant manifolds associated with the UPO. 

As a second quantitative comparison for the single-null tokamak divertor map, we present the computed transient complexity coefficients in Tab.~\ref{tab:BM}, considering all the chosen ensembles of initial conditions, including the special one. We readily see that $c$ can properly differentiate between the transient behaviors observed in Figs.~\ref{fig:BM_profile} and~\ref{fig:BM_special}. For Ens.~1 through 4, the values of $c$ decrease as the ensemble location gets closer to the escape boundary. Moreover, for the special set S, the calculated coefficient accurately expresses how complex, on average, it is the path of an escape orbit in this case.

\begin{table}[!ht]\normalsize
\centering
\caption{Transient complexity coefficient $c$ for the analyzed cases in the single-null tokamak divertor map.\label{tab:BM}}
\begin{tabular}{ c c }
\hline\hline
\hspace*{0.5cm}\textbf{Ensemble}\hspace*{0.5cm}&
\hspace*{0.5cm}\textbf{Coefficient $c$}\hspace*{0.5cm}\\
\hline
\hline
$1$ & $5.723 \times 10^{-1}$\\
\hline
$2$ & $1.344 \times 10^{-1}$\\
\hline
$3$ & $1.408 \times 10^{-2}$\\
\hline
$4$ & $2.908 \times 10^{-3}$\\
\hline
S & $7.539 \times 10^{-1}$ \\
\hline
\end{tabular}
\end{table}

\subsection{Planar Earth-Moon system}

\begin{figure}[!ht]
\centering
\includegraphics[scale=0.83]{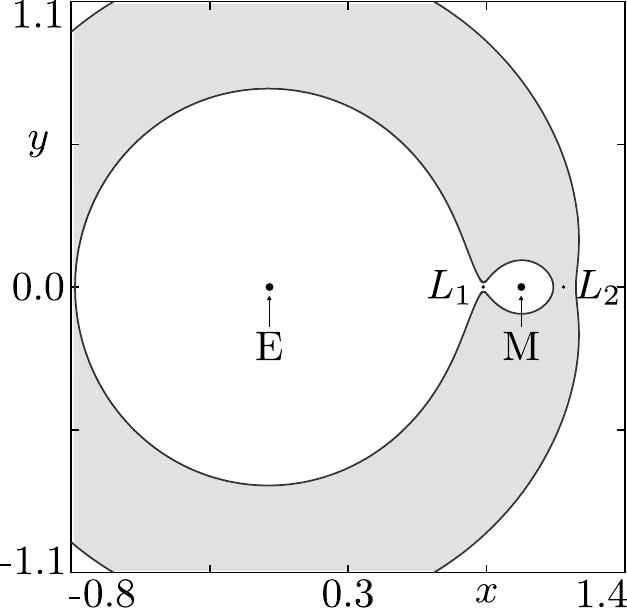}
\caption{The Earth-Moon system as modeled by the planar circular restricted three-body problem. The gray area indicates the forbidden region to which the third particle does not have access for $C=3.187$. Orbits in the vicinity of the Moon can only escape the Moon's realm through the neck in $L_1$.}
\label{fig:3BP_schematics}
\end{figure}

The motion of small bodies in the Earth-Moon system can be modeled, as a first approximation, by the planar circular restricted three-body problem. This model concerns the dynamics of a body with negligible mass under the influence of a two-body gravitational potential \cite{Murray1999}. In a non-inertial reference frame, which rotates with the same constant frequency as the two-body system, the dimensionless equations of motion on the plane $x$-$y$ for the third body are given by
\begin{equation}
\begin{aligned}
\ddot{x}-2\dot{y} &= \dfrac{\partial{\Omega}}{\partial x},\\
\ddot{y}+2\dot{x} &= \dfrac{\partial{\Omega}}{\partial y},\\
\end{aligned}
\label{eq:motion_rotational}
\end{equation}

\noindent with
\begin{equation}
\Omega = \frac{1}{2}(x^2+y^2)+\frac{1-\mu}{r_E}+\frac{\mu}{r_M},
\label{eq:omega}
\end{equation}

\noindent where $\mu=1.215\times10^{-2}$, the ratio between the mass of the Moon and the system's total mass. $r_E$ and $r_M$ are the distances from the primaries, Earth and Moon, which are located at $(-\mu,0)$ and $(1-\mu,0)$, respectively.
The system's schematic is shown in Fig.~\ref{fig:3BP_schematics}.

From Eq.~\eqref{eq:motion_rotational}, we can derive the Jacobi constant of motion $C(x,y)=2\Omega-\dot{x}^2-\dot{y}^2$. It restricts the dynamics of the system to a three-dimensional surface and also delimits the accessible region in coordinate space $x$-$y$. There are two Lagrangian equilibrium points called $L_1$ and $L_2$ next to the Moon and collinear to the primaries. If we set the Jacobi constant between the values of $C$ for these points, namely, $C_{L_1}\approx3.188$ and $C_{L_2}\approx3.172$, we arrive at a situation where orbits that start near the Moon can transfer to the Earth's vicinity but cannot leave the system.

\begin{figure*}[!ht]
\centering
\includegraphics[scale=0.83]{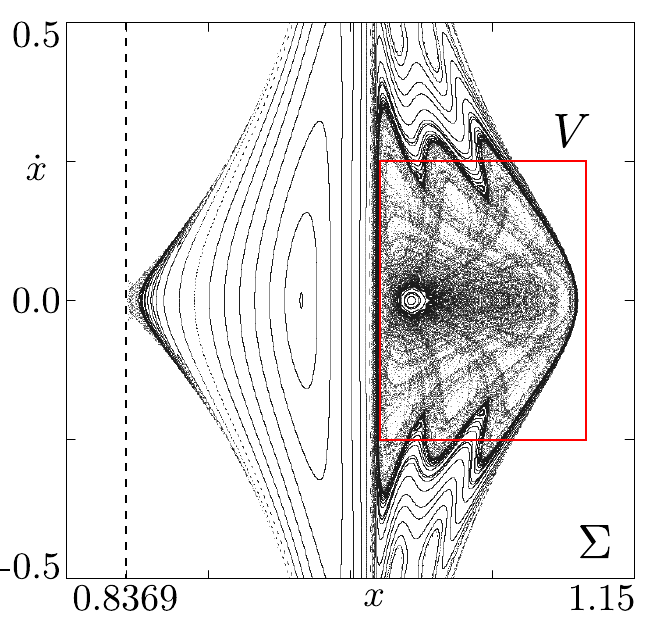}
\includegraphics[scale=0.83]{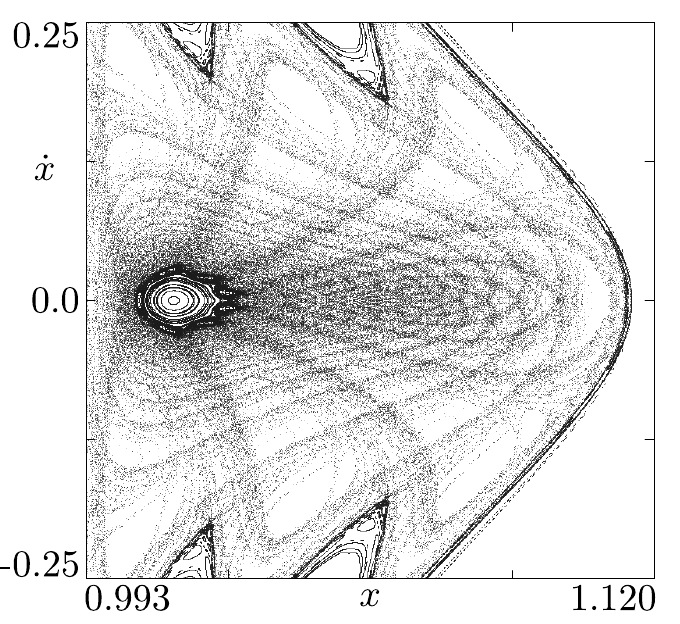}
\caption{Phase space $x$-$\dot{x}$ for the planar Earth-Moon system calculated at the surface of section $\Sigma$. (Left) Full phase space, where the dashed line marks the escape threshold. (Right) Zoom-in on our region of interest $V$.}
\label{fig:3BP_phase_space}
\end{figure*}

\begin{figure*}[!ht]
\centering
\includegraphics[scale=0.83]{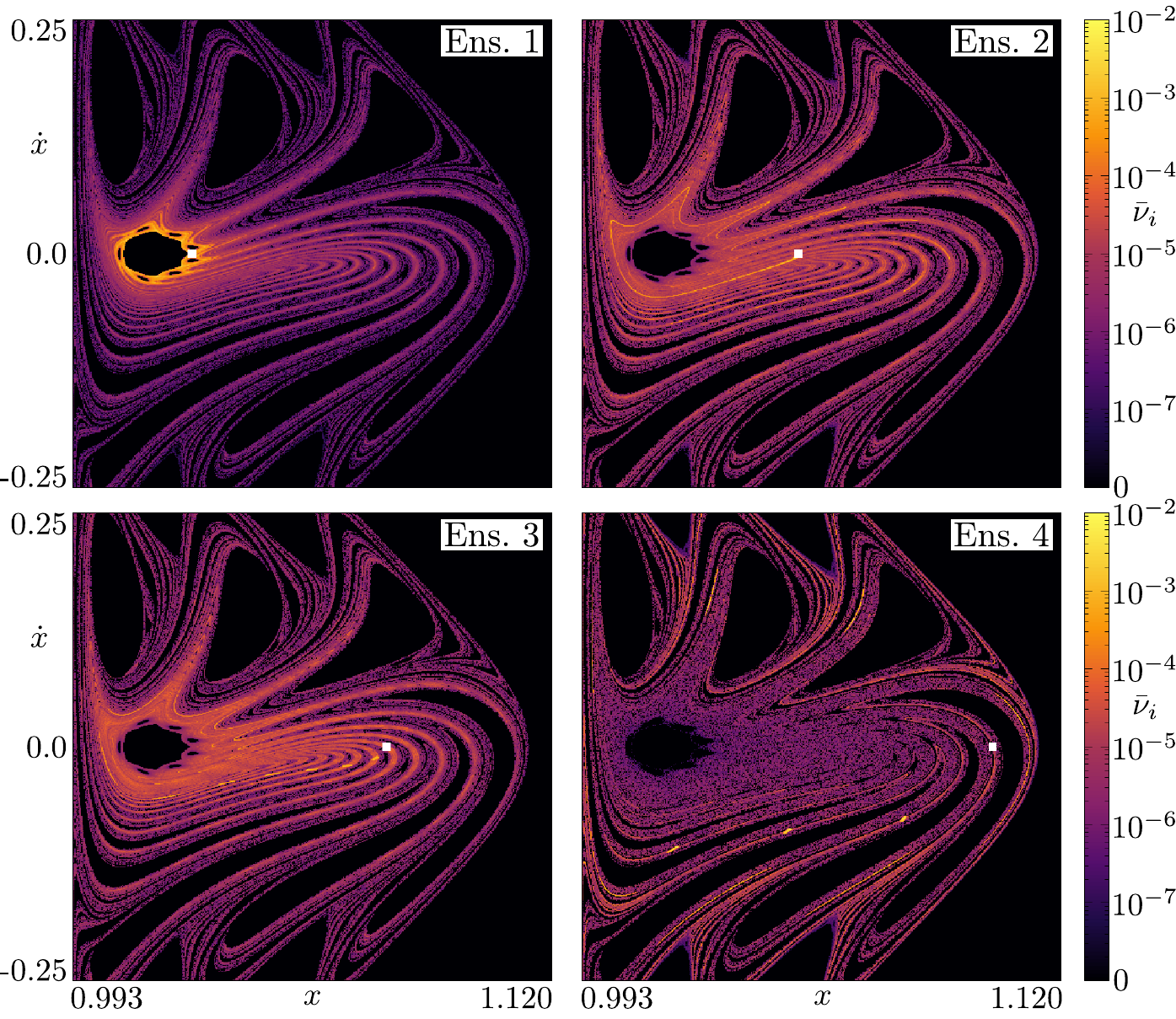}
\caption{Profiles of the mean transient measure $\bar{\nu}_i$ in logarithmic scale for the Earth-Moon system, calculated on a $512\times512$ grid in the region $V$ of phase space $x$-$\dot{x}$. The ensembles of initial conditions are chosen in the surface of section $\Sigma$ and are represented by the small white squares, which are not in scale.}
\label{fig:3BP_profile}
\end{figure*}

In this work, we set $C=3.187$ and we consider that orbits escape when they exit the Moon's realm and enter the Earth's one, which are separated by $L_1$. Therefore, the escape condition is given by $x<x_{L_1}$. The equations of motion are numerically integrated up to $T_{max}=5\times10^{3}$ units of time using the explicit embedded Runge-Kutta Prince-Dormand 8(9) method \cite{Galassi2001} and the orbits are analyzed on a surface of section $\Sigma$ defined by
\begin{equation}
    \Sigma = \{ (x,y,\dot{x},\dot{y}) ~|~ x_{L_1}<x<x_{L_2},~y=0,~\dot{y}>0 \},
\end{equation}

\noindent where $x_{L_1}\approx0.8369$ and $x_{L_2}\approx1.1556$ are the location on the $x$-axis of $L_1$ and $L_2$, respectively.

Figure~\ref{fig:3BP_phase_space} presents the system's phase space $x$-$\dot{x}$ in our surface of section $\Sigma$, along with the escape threshold $x=x_{L_1}$ and the region $V\subset \Sigma$ that we are interested in. We can observe one main stability region formed by regular solutions together with a large chaotic sea. There is a clear presence of stickiness around the stability region and also areas with a higher or a lower density of orbits in the chaotic sea.

It is important to note here that the method presented in Sec.~\ref{sec:concepts} does not require the escape threshold to be inside the region of interest. In this case, $V$ is far from the neck region that divides the realms and it contains the direct periodic orbit around the Moon for this Jacobi constant \cite{Restrepo2018}.

\begin{figure*}[!ht]
\centering
\includegraphics[scale=0.83]{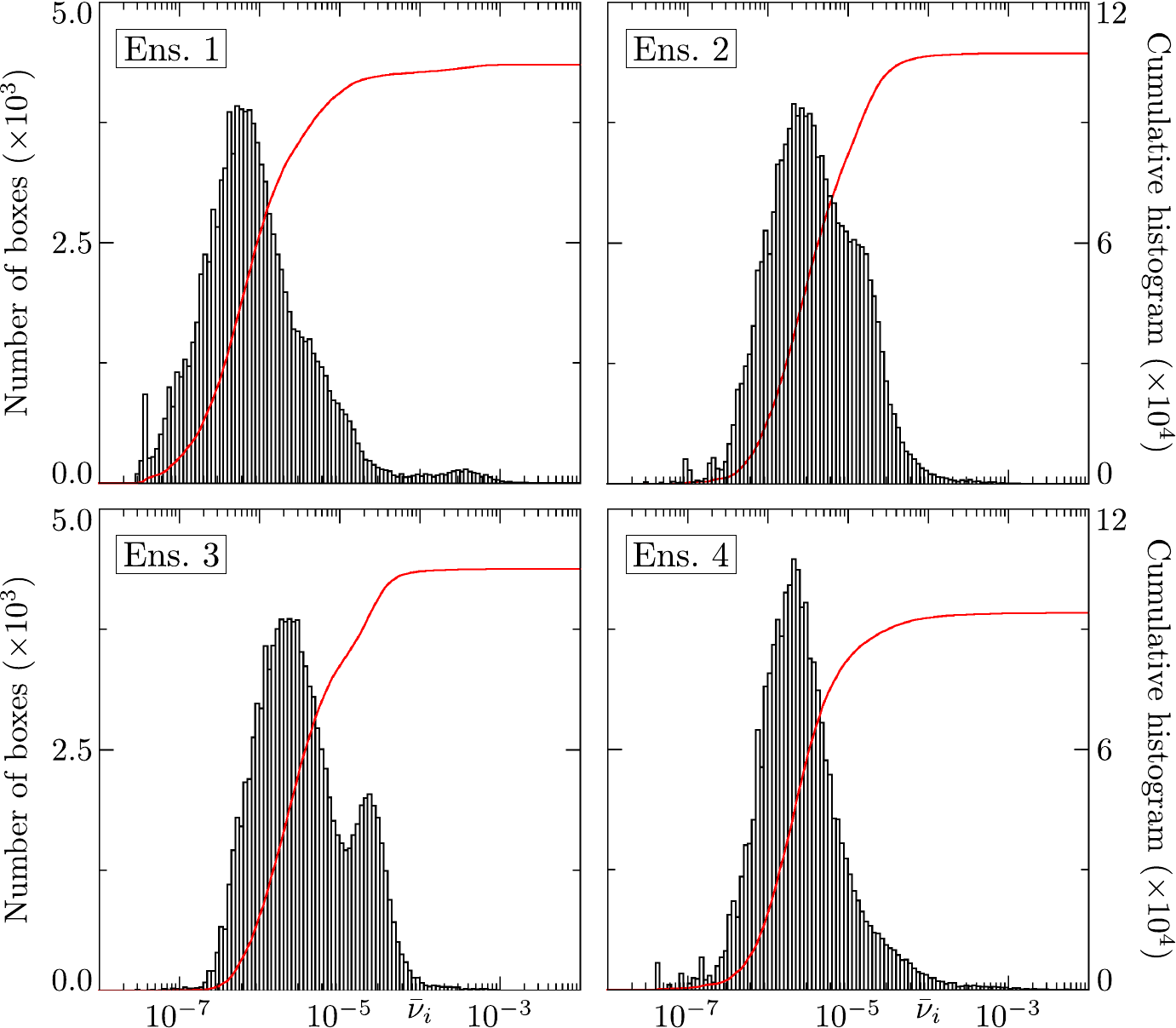}
\caption{Histogram (black rectangles) and cumulative histogram (red line) of the mean transient measure $\bar{\nu}_i$ for the Earth-Moon system. The grid is formed by $512^2=26.2144\times10^{4}$ boxes and only the ones visited by an orbit are considered.}
\label{fig:3BP_histogram}
\end{figure*}

\begin{figure*}[!ht]
\centering
\includegraphics[scale=0.83]{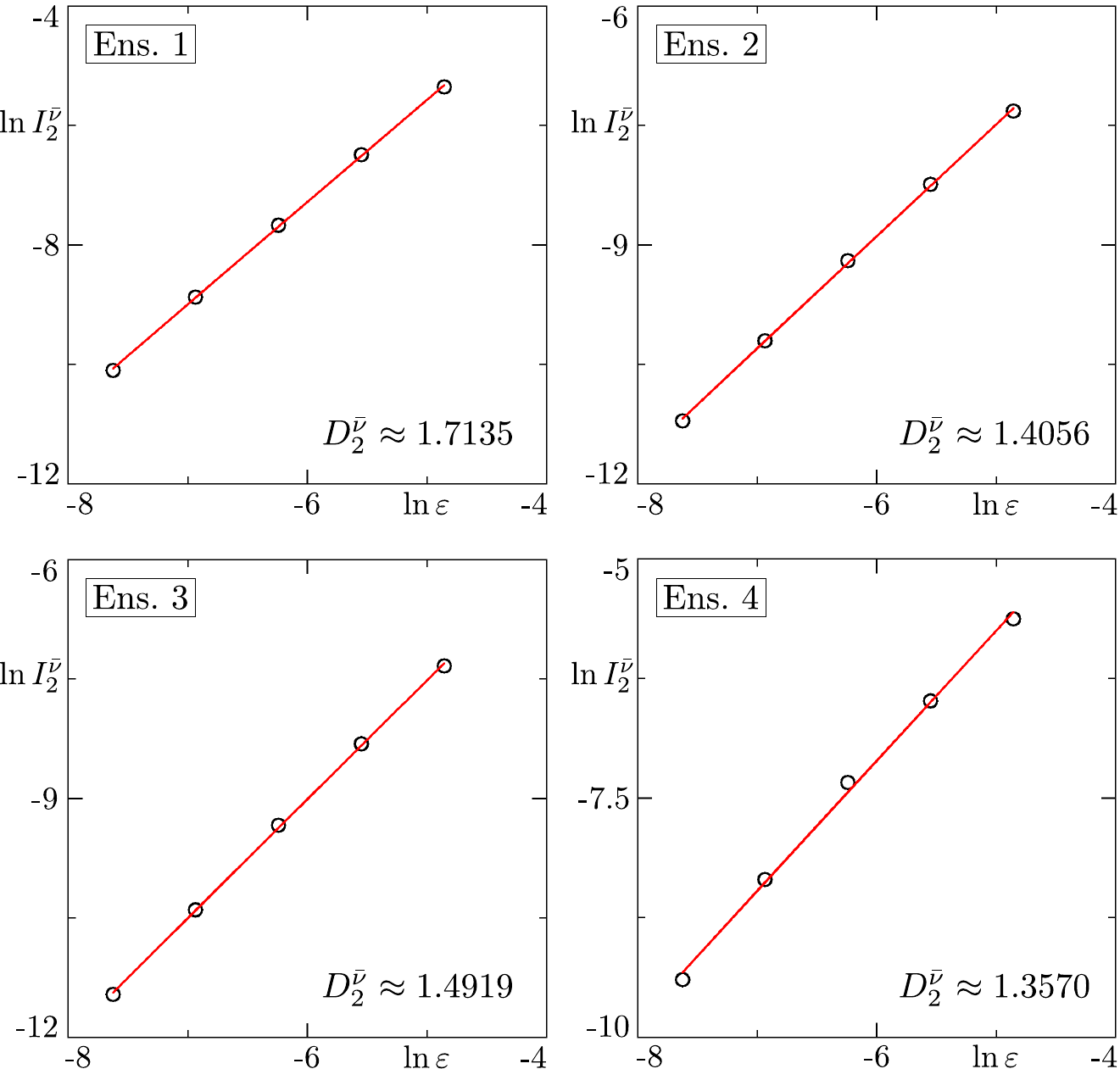}
\caption{Linear fitting of $\ln{I^{\bar{\nu}}_2}$ as a function of $\ln{\varepsilon}$ for the Earth-Moon system, where $D^{\bar{\nu}}_2$ is the angular coefficient. The values of the box side-lengths are $\varepsilon=1/128$, $1/256$, $1/512$, $1/1024$ and $1/2048$ for a phase space normalized to the unity square.}
\label{fig:3BP_dimension}
\end{figure*}

In order to investigate this system, we choose our four ensembles of $M=10^{4}$ initial conditions in the region $V$. These ensembles are now formed by rectangles of size $5\times10^{-4}$ by $4\times10^{-3}$ in the phase space $x$-$\dot{x}$ and are equally distant from each other, with Ens. 1 centered at approximately $(1.025,~0.0)$ and Ens. $4$ at $(1.102,~0.0)$.
In Fig.~\ref{fig:3BP_profile}, we present the mean transient measure profiles for a grid of $512\times512$ boxes.

Ensemble $1$ is chosen in the neighborhood of the stability region, as we did for the special case in the tokamak system. In this system, however, we do not have a series of island chains, but rather one main stability region along with a large chaotic sea. As was the case there, we readily notice a higher visitation frequency in the boxes around the stability region, highlighting the stickiness effect and also delineating the invariant manifolds associated with the period-7 UPO in which the ensemble is centered. 
It is interesting to observe, though, that none of the other cases experience the same stickiness effect. While the orbits that begin in Ens. 2 and 3 spread across the region $V$ with a higher visitation frequency in the middle section, the ones that begin in Ens. 4 seem to concentrate more on the outer part. Therefore, we observe here three very distinct transient behaviors.

In Fig.~\ref{fig:3BP_histogram}, we present both the histogram and the cumulative histogram of the mean transient measure for all cases, considering only the visited boxes.
As expected due to the stickiness effect, Ensemble~$1$ leads to the highest number of boxes with high mean transient measures, visible as a small bump in the first histogram. Furthermore, even though Ens. 2 and 3 look similar in Fig.~\ref{fig:3BP_profile}, they present different distributions, with the latter having two clear peaks. This could indicate the presence of another UPO in the system, which would be influencing the path of these orbits. As for Ens. 4, the distribution is thinner than the others and the reason why this is so is addressed later.

As before, we initially quantify our observations by calculating the transient correlation dimension for all cases. In Fig.~\ref{fig:3BP_dimension}, we show $I^{\bar{\nu}}_2$ as a function of the box side-length $\varepsilon$ in a log-log plot, along with the linear fitting for each case and the corresponding value for $D_2^{\bar{\nu}}$. The results obtained again validate Eq.~\eqref{eq:escape_correlation_dimension_def} and show that such quantity is well suited for characterizing the transient behavior in this system as well.
The calculated dimension is higher for Ens. $1$ and lower for Ens. $4$, as we would expect by looking at Fig.~\ref{fig:3BP_profile}. 
In here, however, the relation between transient correlation dimension and distance from the escape threshold is not linear, since $D_2^{\bar{\nu}}$ is slightly higher for Ens. $3$ when comparing to Ens. $2$, which suggests a more complicated transient scenario for this system.

There are two observations we need to make about the interpolation for Ens. $4$. First, the total number of visited boxes is significantly lower than the other cases, which is the main information we can extract from the cumulative histograms in Fig.~\ref{fig:3BP_histogram}. Second, there is a lower number of orbits with high $\bar{\nu}_i$, as we can also see from the histograms. Therefore, the statistics necessary for calculating $I_2^{\bar{\nu}}$ and, consequently, the escape dimension correlation,0 are not optimal in this case.

\begin{figure}[!ht]
\centering
\includegraphics[scale=0.83]{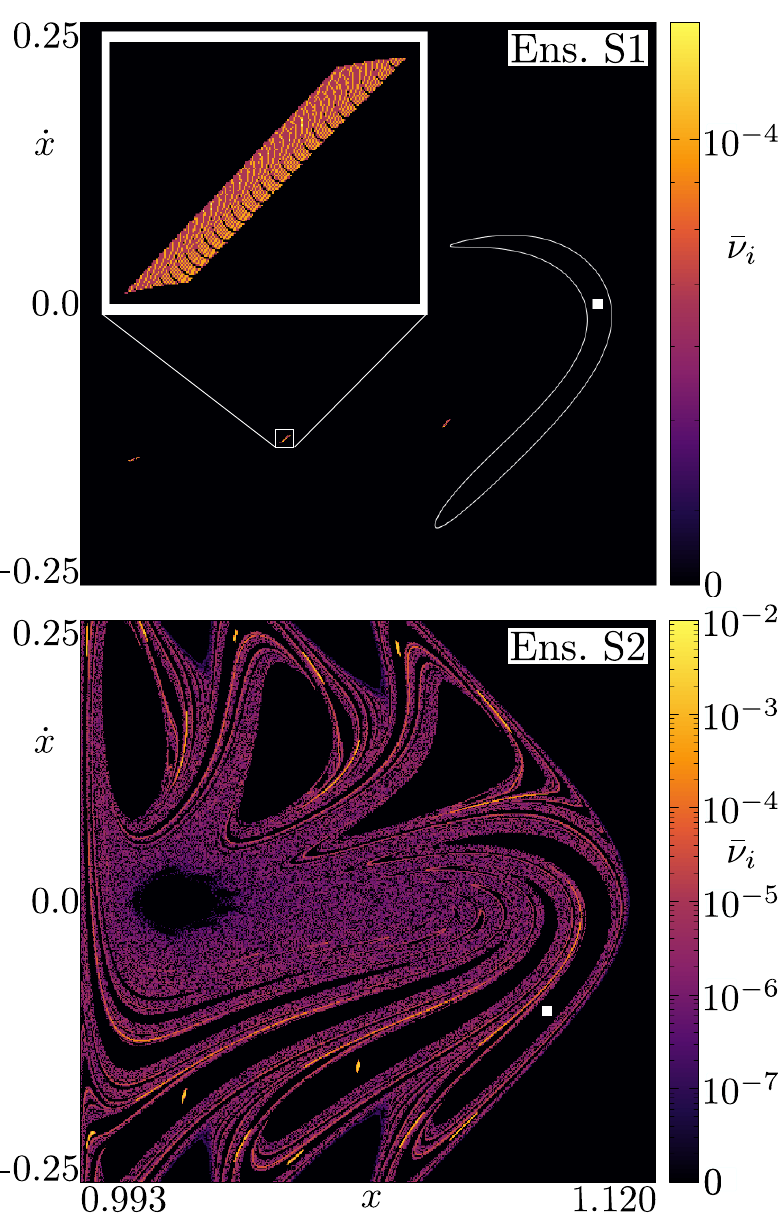}
\caption{Profiles of the mean transient measure in logarithmic scale for the special cases in the Earth-Moon system. The ensembles are represented out of scale by the white squares. For S1, we also show the contour of the chosen black lake in white and the magnification of the profile for the second crossing.}
\label{fig:3BP_special}
\end{figure}

Another information we obtain from the cumulative histograms is that less than half of the grid boxes are being visited in all cases, which can be explained in the following manner. From Fig.~\ref{fig:3BP_profile}, we observe the presence of ``black lakes", which are not visited in any of the analyzed cases and yet are not composed by regular structures (see Fig.~\ref{fig:3BP_phase_space}). As special cases, then, we choose two ensembles of initial conditions inside one of these regions and calculate the respective mean transient measure profiles. The results are presented in Fig.~\ref{fig:3BP_special}, with ensemble S1 centered at $(1.107,0.0)$ and S2 at $(1.096,-0.09744)$, approximately.

For Ens. S1, we observe an interesting situation where all orbits evolve closely together and exit the system after crossing $V$ three times, which means that there are fast escape routes inside these regions.
However, for Ens. S2, the orbits do not rapidly escape and, instead, they spread across the phase space after crossing $V$ through other black lakes. This means that it is possible for orbits that begin in one of these regions to access the same part of the phase space that was visited in the previous cases, though the opposite is not true as we can see from Fig.~\ref{fig:3BP_profile}.

The black lakes are actually formed by the crossing with the surface of section of the unstable manifolds of the Lyapunov orbit, an UPO that revolves around the equilibrium point $L_1$. These geometrical structures are two-dimensional surfaces and are responsible for the transport of orbits that enter the Moon's realm \cite{Ozorio1990,Perozzi2010}. 
Ensemble S1 is fully inside an intersection between manifolds of different stabilities, which causes the orbits that begin in this set to exit the Moon's vicinity following the stable one \cite{deOliveira2020CMDA}. Furthermore, Ensemble 4 is partially inside an intersection and, consequently, a portion of the orbits originating in this set rapidly escape the system, lowering the number of visited boxes, as seen in the cumulative histogram of Fig.~\ref{fig:3BP_histogram}. 

We now proceed to the calculation of the transient complexity coefficient, which is shown in Tab.~\ref{tab:3BP} for all the cases analyzed in this system, including the two special ones.
For Ensembles 1 through 4, the coefficient $c$ differentiates the transient behavior between the cases similar to the transient correlation dimension, with the value for Ens.~3 being higher than for Ens.~2.
For Ens.~S1, in particular, the transient complexity coefficient takes into consideration the system's fast dynamics and correctly gives a value lower than the regular cases. 
\begin{table}[!ht]\normalsize
\centering
\caption{Transient complexity coefficient $c$ for the analyzed cases in the planar Earth-Moon system.\label{tab:3BP}}
\begin{tabular}{ c c }
\hline\hline
\hspace*{0.5cm}\textbf{Ensemble}\hspace*{0.5cm}&
\hspace*{0.5cm}\textbf{Coefficient $c$}\hspace*{0.5cm}\\
\hline
\hline
$1$ & $1.718 \times 10^{-1}$\\
\hline
$2$ & $1.569 \times 10^{-3}$\\
\hline
$3$ & $2.377 \times 10^{-3}$\\
\hline
$4$ & $9.460 \times 10^{-4}$\\
\hline
S1 & $4.483 \times 10^{-5}$\\
\hline
S2 & $1.235 \times 10^{-3}$\\
\hline
\end{tabular}
\end{table}

\section{Conclusions}
\label{sec:conclusions}

In this work, we introduced a practical numerical method that visually illustrates and quantifies the transient behavior of Hamiltonian systems with a defined escape, and we investigated the dynamics of two physical systems with very different dynamical scenarios: a tokamak with a single-null divertor and the planar Earth-Moon system. The first one was described by a two-dimensional map, which presented a complex structure of island chains and associated unstable periodic orbits of varied periods. The second one was modeled by a four-dimensional time-continuous system with a constant of motion and presented a phase space structure composed of one main regular region along with a large chaotic sea.

By plotting the profiles of the mean transient measure for different ensembles on these systems, we verified that, depending on the location of the ensemble of initial conditions, the escape orbits experienced very different paths in the phase space before reaching the escape condition, which characterized distinct transient scenarios. Furthermore, the profiles provided a clear picture of the stickiness phenomenon and highlighted the influence of particular invariant manifolds. Later, with the mean transient measure histograms, we were able to show the differences between the phase space distributions associated with each ensemble, and also between the two physical systems.

The transient scenarios were quantified by two distinct parameters, the transient correlation dimension, and the transient complexity coefficient, both of which were capable of determining which situations lead to the most complex behavior. The transient correlation dimension was defined directly from the mean transient measure, without further considerations. This quantity does not dependent on the box side-length, since it is defined in the limit $\varepsilon\to0$, which makes it somewhat more general. However, calculating the transient correlation dimension was costly and it required a high number of box visitation.

The transient complexity coefficient, on the other hand, takes into consideration how many orbits of the ensemble go through a given box of the grid and how fast said orbits exit the system. These dynamical aspects are then weighted in on the mean transient measure over-the-grid summation. As a result, the transient complexity coefficient returned numerical values on a non-linear scale, providing a clear distinction between the analyzed cases. On the downside, this quantity is related to a specific grid box side-length, which needed to be properly chosen.

In summary, we showed that the mean transient measure is an effective numerical tool for visually describing and characterizing the different transient scenarios that may arise in Hamiltonian systems.

\section*{Acknowledgements}
This study was financed in part by the Coordena\c c\~ao de Aperfeiçoamento de Pessoal de Nível Superior - Brasil (CAPES) - Finance Code 001, the São Paulo Research Foundation (FAPESP, Brazil), under Grants No.  2018/03211-6 and 2018/03000-5, and the Brazilian Federal Agency CNPq, under Grant No. 302665/2017-0. 

\section*{CRediT authorship contribution statement}

\textbf{Vitor M. de Oliveira:} Conceptualization. Methodology. Software. Investigation.  Writing - Original Draft \textbf{Matheus S. Palmero:} Software. Validation. Investigation. Writing - Original Draft. \textbf{Iberê L. Caldas:} Writing - Review \& Editing. Supervision. Funding acquisition.

 \bibliographystyle{elsarticle-num} 
 \bibliography{natural}





\end{document}